\documentclass[reprint,showpacs,aps,prl,10pt,superscriptaddress,]{revtex4-1}
\usepackage{amsmath, amsfonts, amssymb,mathtools,mathrsfs}
\usepackage{graphicx}
\usepackage[english]{babel}
\usepackage{color}

  \makeatletter
    \renewcommand\@make@capt@title[2]{%
     \@ifx@empty\float@link{\@firstofone}{\expandafter\href\expandafter{\float@link}}%
      {\textbf{#1}}\@caption@fignum@sep#2\quad}%
    \makeatother
   
\makeatletter 
\renewcommand{\fnum@figure}{\textbf{Figure~\thefigure}}
\makeatother

\newcommand{\beginsupplement}{
        \renewcommand{\thetable}{S\arabic{table}}
        \renewcommand{\thefigure}{S\arabic{figure}}
	 \renewcommand{\theequation}{S\arabic{equation}}
	 \renewcommand{\thesection}{S\arabic{section}}	 
     }

\DeclarePairedDelimiterX{\norm}[1]{\lVert}{\rVert}{#1}

\begin{document}

\title{Entropy, Ergodicity and Stem Cell Multipotency}

\author{Sonya J. Ridden}
\affiliation{Mathematical Sciences, University of Southampton, SO17 1BJ, UK}
\author{Hannah H. Chang}
\affiliation{5AM Ventures, Boston, MA 02109, USA}
\author{Konstantinos C. Zygalakis}
\affiliation{Mathematical Sciences, University of Southampton, SO17 1BJ, UK}
\author{Ben D. MacArthur}
\email{Correspondence to bdm@soton.ac.uk}
\affiliation{Mathematical Sciences, University of Southampton, SO17 1BJ, UK}
\affiliation{Centre for Human Development, Stem Cells and Regeneration, University of Southampton, SO16 6YD, UK}

\date{\today}

\begin{abstract}
Populations of mammalian stem cells commonly exhibit considerable cell-cell variability. However, the functional role of this diversity is unclear. Here, we analyze expression fluctuations of the stem cell surface marker Sca1 in mouse hematopoietic progenitor cells using a simple stochastic model and find that the observed dynamics naturally lie close to a critical state, thereby producing a diverse population that is able to respond rapidly to environmental changes. We propose an information-theoretic interpretation of these results that views cellular multipotency as an instance of maximum entropy statistical inference. 
\end{abstract}

\pacs{87.10.Vg, 89.70.-a, 89.70.Cf, 87.18.Tt, 87.10.Mn, 87.17.Aa}

\maketitle 

Clonal populations of unicellular organisms often exhibit phenotypic diversity, which confers selective advantage under adverse environmental conditions. Well-known examples include antibiotic bacterial persistence, the lysis-lysogeny switch of $\lambda$-phage, competence development and sporulation of \emph{B. subtilis}, and lactose uptake by \emph{E. coli} \cite{Veening, *Veening2}.  The ubiquity of this phenomenon indicates that it is a generic, evolvable, mechanism that facilitates collective cellular dynamics by enabling robust, rapid responses to diverse environmental changes. Recently, stochastic fluctuations in expression of important marker proteins have been seen to generate functional diversity within multipotent mammalian stem cell populations, suggesting a similar role for cell-cell variability in higher organisms \cite{Graf, *Cahan,*kalmar2009regulated}. These observations have motivated speculation that functional multipotency (the ability to differentiate along a number of distinct cellular lineages) is a collective property of stem and progenitor cell populations, reflective of fitness constraints imposed at the population, rather than individual cell, level \cite{lei, *MacArthur, *garcia2012towards}. This perspective is appealing since such regulated cell-cell variability in principle allows a cellular population to remain primed to respond quickly to a range of different differentiation cues while remaining robust to cell loss.  However, convincing demonstrations of the potency of individual stem cells appear to argue strongly against such a collective view (for example, single long-term repopulating hematopoietic stem cells are able to fully reconstitute the blood system of lethally irradiated adult mice and small numbers of pluripotent stem cells are able to rescue development of genetically compromised embryos \cite{Osawa, *wang}). Thus, it is still unclear how population-level and cell-intrinsic regulatory programs interact to control mammalian stem and progenitor cell dynamics. 
 
Here we propose  a theoretical framework that reconciles these disparate observations, which views cellular multipotency as an instance of maximum entropy statistical inference. In this view, individual cells satisfy any minimal regulatory constraints imposed upon them (such as basic metabolic requirements, etc.) yet, in the absence of defined instructions, are maximally noncommittal with respect to their remaining molecular identity, thereby generating a diverse population that is able to respond optimally to a range of unforeseen future environmental changes. Thus, rather than viewing the multipotent cell state as an attractor of the underlying molecular regulatory dynamics (i.e. associating cellular identities with well-defined, stable, patterns of gene expression -- a common modeling assumption, that has received some experimental validation for differentiated cell types \cite{huang2005cell}), individual multipotent cells are characterized by fundamental \emph{uncertainty} in their molecular state and their populations exhibit variability in accordance with this intrinsic uncertainty. However, since this model exchanges the attractor hypothesis at the single cell level for an ergodicity assumption for the underlying stochastic processes, each individual cell has the latent potential to assume every identity within the population, and thereby retains the regenerative capacity of the entire population. As this view is fundamentally stochastic, its corollary is that regulation of multipotency occurs at the level of probabilities (i.e. at the population level), rather than at the individual cell level. 

\begin{figure}[t!] 
\includegraphics[width=0.475\textwidth]{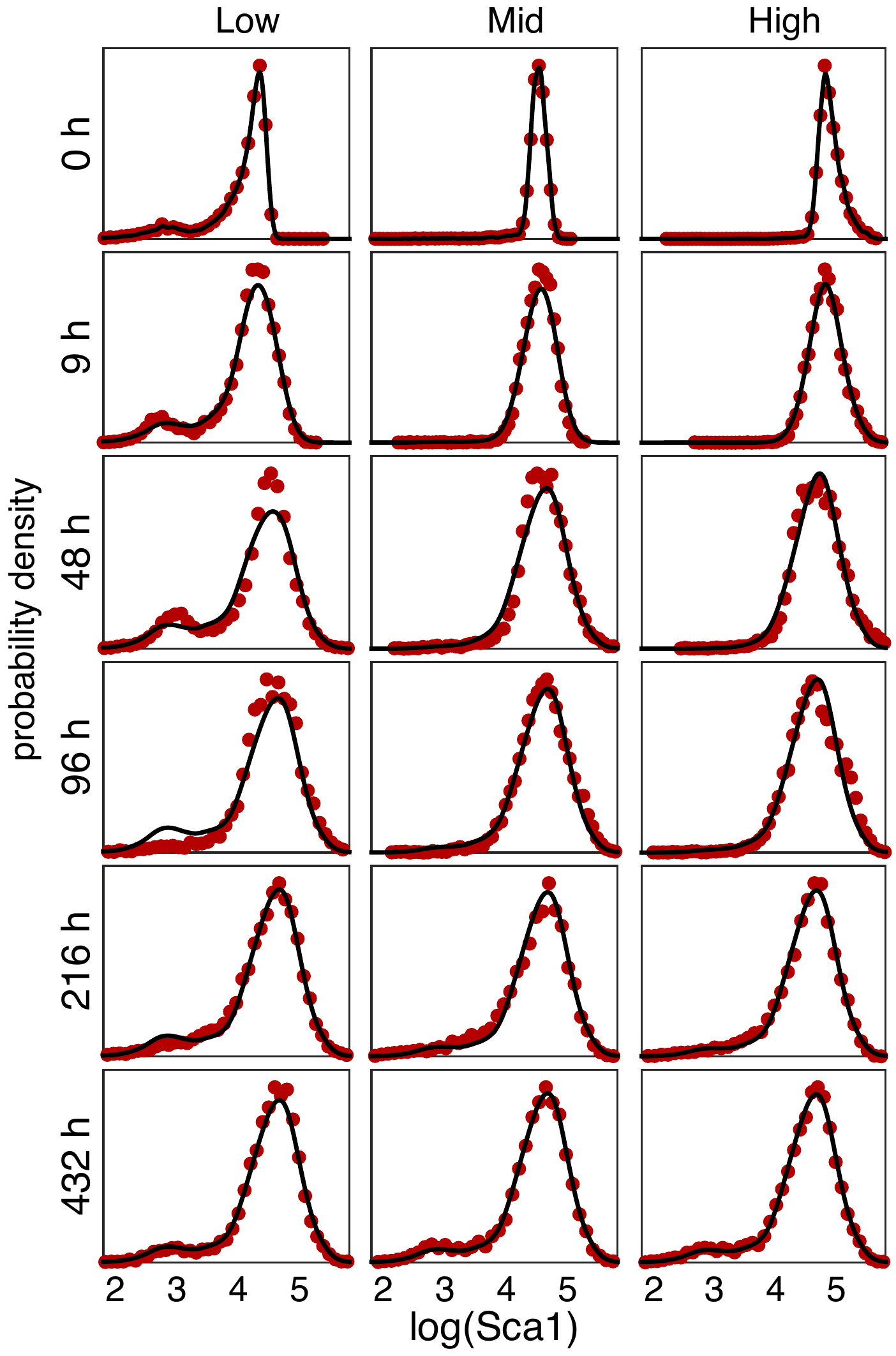}  
\caption{Model fit to experimental data. Model simulations using the same estimates of $\psi(x)$ and $\sigma$ are shown against the three independent experimental time-series; simulations differ only in the experimentally prescribed initial conditions. Data is in dark red and the fitted model is in black. The potential $\psi (x)$ was estimated numerically via Eq.~\eqref{bg} using aggregated data from the final time point. Color online.}
\label{FIG:evol}
\end{figure}

In order to illustrate this perspective we consider here the expression dynamics of the stem cell surface marker Sca1 (stem cell antigen 1) in populations of multipotent \textsc{eml} mouse hematopoietic progenitor cells. It has previously been shown that Sca1 levels fluctuate stochastically in \textsc{eml} cells in culture, with extrinsic `transcriptome-wide' noise driving transitions between Sca1 high and Sca1 low states, which transiently prime individual cells for erythroid and myeloid differentiation respectively and generate a characteristically bimodal Sca1 expression distribution within the population (see Fig. \ref{FIG:evol}, bottom panel and Ref. \cite{chang}). However, the underlying mechanisms by which these stochastic fluctuations are regulated are not known. In the absence of this knowledge we assume here that the intracellular dynamics of the Sca1 expression level $z(t)$ are described by a generic stochastic differential equation:
\begin{equation} \label{sde}
\frac{\textrm{d}z}{\textrm{d}t}=a(z)+\sqrt{2d(z)}\xi(t), \nonumber
\end{equation}
where $\xi(t)$ is a standard one-dimensional white noise process [$\left \langle \xi(t) \right \rangle=0$ and $\left \langle \xi(t)\xi(s) \right \rangle =\delta(t-s)$] and $d(z)$ accounts for fluctuations in Sca1 levels due to both intrinsic sources (i.e. noise in the molecular processes involved in Sca1 production/decay, such as transcription, translation, translocation and degradation etc.) and extrinsic sources (i.e. fluctuations in upstream regulators and uncontrolled environmental noise). Rather than model Sca1 levels directly it is convenient to introduce a reaction coordinate $x(z)$ such that the Fokker-Planck equation for the probability density $\rho(x,t)$ has the form
\begin{equation} \label{fp_real}
\frac{\partial \rho}{\partial t} =  L(\rho), \quad
L(\rho) = \frac{\partial}{\partial x} \left ( \frac{\mathrm{d} \psi}{\mathrm{d} x} \, \rho \right )  + \sigma \frac{\partial^2 \rho}{\partial x^2},
\end{equation}
with scalar potential $\psi(x)$ and diffusion coefficient $\sigma$. Such a transformation, which maps the original dynamics to those of a Brownian particle in a one-dimensional potential field, may be achieved by application of It\={o}'s lemma (see Supplemental Material for details). The stationary solution of  Eq.~\eqref{fp_real} is the Boltzmann-Gibbs distribution
\begin{equation} \label{bg}
\rho_\infty (x) = Z^{-1} \exp (-\psi/\sigma ), \;\;\;\;  Z = \int \exp (- \psi/\sigma ) \, \mathrm{d}x. 
\end{equation}
This solution exists so long as $\psi (x)$ grows sufficiently rapidly as $| x | \to \infty$ that the partition function $Z$ remains finite. In this case, the dynamics are ergodic and the free energy
\begin{eqnarray} 
F(\rho) &=& \int \psi \rho \, \mathrm{d}x + \sigma \int \rho \log \rho \, \mathrm{d}x,  \nonumber \\ 
&=& E(\rho) - \sigma S(\rho), \nonumber
\end{eqnarray}
where $E(\rho)$ and $S(\rho)$ are the energy and entropy functionals respectively, is a Lyapunov functional for the dynamics. Thus, in order to model Sca1 dynamics phenomenologically we need only chose an appropriate reaction coordinate $x$ and form for the potential $\psi(x)$. 

Since noise in protein expression often scales with abundance, a natural choice for the reaction coordinate is $x=\log{z}$, as has been taken elsewhere (see Supplemental Material for details) \cite{Bar-Even,*sisan2012predicting}. In the absence of detailed information on how Sca1 fluctuations are regulated, the potential $\psi(x)$ may be estimated numerically from the empirical Sca1 distribution by inverting Eq.~\eqref{bg}. The model then has a single free parameter, the diffusion coefficient $\sigma$, which sets the timescale for the dynamics. 

\begin{figure}[t!]
\centering
\includegraphics[width=.475\textwidth]{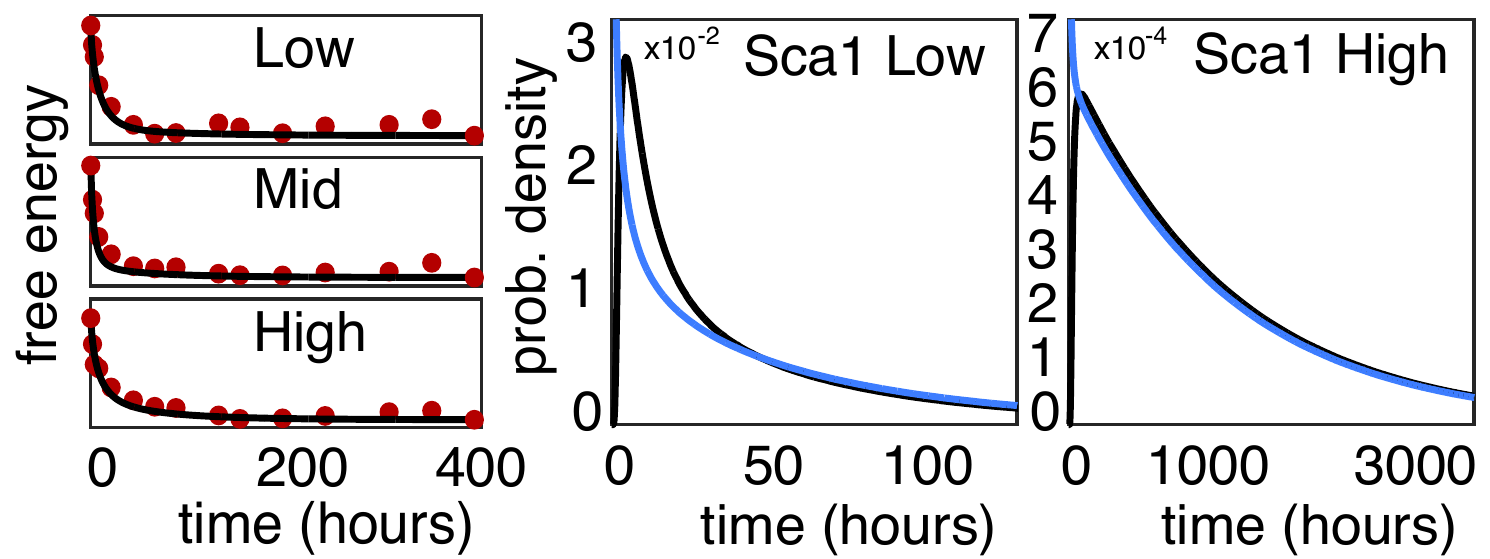}
\caption{(\emph{Left}) Convergence to equilibrium with respect to the free energy. Exponential convergence was observed from all three initial conditions for large time, in accordance with Eq.~\eqref{fp_real}. First passage time (\textsc{fpt}) distributions in the Sca1 low (\emph{Middle}) and high (\emph{Right}) states. The \textsc{fpt} distributions $F_{X}(x_X,t)$ starting at the local minima of the potential $\psi(x)$ are shown in black; the expected \textsc{fpt} distributions $\langle F_X \rangle (t)$ averaging over all initial conditions in $X \in \{L, H \}$ are shown in blue. Color online.}
\label{FIG:res}
\end{figure}

Estimates of $\sigma$ and $\psi(x)$ were obtained by model fitting using maximum likelihood estimation to evolving Sca1 expression distributions obtained experimentally using flow cytometry starting from pre-selected populations of Sca1 low, mid, and high expressing cells as they equilibrate in culture over a period of 18 days (obtained in Ref. \cite{chang}). Despite the simplicity of this model, an excellent agreement with the experimental time-series data was observed from all three initial conditions, using the same numerically estimated potential and the same estimate of $\sigma$ (Figs. \ref{FIG:evol}--\ref{FIG:res}).  

It has previously been argued, based upon analysis of changing proportions of cells in the Sca1 high and low states, that the observed dynamics are characterized by slow `sigmoidal' relaxation towards the stationary state \cite{chang}. Since a constant probability flux across a barrier naturally leads to exponential relaxation, it was suggested that these dynamics indicate deviation from expected first-order kinetics, possibly due to regulation of Sca1 fluctuations by cell-cell communication or autocrine signaling. However, it is apparent that such recourse is not needed since in all 3 cases the experimental system is initially far from equilibrium, and therefore far from the regime in which first-order kinetics apply. Rather, in accordance with standard reaction-rate theory, the dynamics are characterized by an initial transient period during which local equilibrium is first established within each potential well, before transitions between wells occur \cite{hanggi1990reaction}. Examination of the free energy (which is a natural way to assess convergence to equilibrium \cite{markowich2000trend, *jordan1998variational}) shows that this separation of timescales naturally generates the observed convergence dynamics without the need to include additional regulatory mechanisms in the model (see Fig. \ref{FIG:res}, left). These results indicate that the observed Sca1 expression dynamics are well described by a simple ergodic process in which individual cells behave independently with respect to Sca1 fluctuations.

This ergodic property is useful since it allows inference of the behavior of individual cells from the population dynamics. While stochastic excursions into the Sca1 high and low states have previously been seen to transiently confer different lineage biases to individual progenitor cells in culture, the timescales upon which these excursions occur at the single cell level are not known. Thus, the distribution of first passage times (\textsc{fpt}s) out of the Sca1 low and high states are of particular interest. Defining the ranges of Sca1 low and high expression as $L = (-\infty,x_{0})$ and $H = (x_{0},\infty)$ respectively, where $x_0$ is the intermediate maxima in $\psi(x)$, the \textsc{fpt} $T(x)$ out of $X$ for a cell initially at $x \in X$ (where $X \in \{L,H \}$) may be obtained from the backward Fokker-Planck equation associated with Eq.~\eqref{fp_real}. Denoting $G(x,t)=P(T(x) \geq t)$ we solve:
\begin{equation} \label{bvp1}
\frac{\partial G}{\partial t} = -\frac{\mathrm{d} \psi}{\mathrm{d} x}\, \frac{\partial G}{\partial x}  + \sigma \frac{\partial^2 G}{\partial x^2}, \nonumber
\end{equation}
with initial conditions $G(x,0) = 1$ for $x \in X$ and boundary conditions $G(x_0,t) = \partial G/\partial x(\pm \infty,t) = 0$, from which the \textsc{fpt} distributions $F_{X}(x,t)=-\partial  G / \partial t$ for $X \in \{L,H\}$ may be obtained. Conventionally, the \textsc{fpt} distribution $F_{X}(x,t)$ is evaluated from the local minima $x_X$ of $\psi(x)$ in $X$, since this is the state of highest probability. Alternatively, we can weight each initial position within $X$ according to the probability that the cell is at this position at equilibrium. We thus define the expected \textsc{fpt} distribution with respect to the Gibbs measure, 
\begin{equation}
\langle F_{X} \rangle (t) = \int_{x \in X} \frac{\rho_\infty(x)}{w_{X}} F_{X}(x,t) \, \textrm{d}x, \nonumber
\end{equation} 
where $w_X = \int_{x \in X} \rho_\infty(x) \textrm{d}x \in [0,1]$ is the weight of the population in $X$. Numerical approximations to $F_{X}(x_X,t)$ and $\langle F_X \rangle (t)$ are shown in Fig. \ref{FIG:res}. These distributions yield mean \textsc{fpt}s of 60/56 hours for the low state and 1573/1487 hours for the high state using $F_{X}(x_X,t)$ and $\langle F_X \rangle (t)$ respectively. These timescales are substantially longer than the \textsc{eml} cell cycle time (approx. 10\,--\,14 hours \cite{bowie, *nygren}),  and therefore suggest that Sca1 fluctuations are not simply a consequence of the cell-cycle. Rather, by setting the expected length of time that a pair of cells initially at the same position (e.g. daughter cells from the same cell division) will forget their common origin  -- and therefore the expected length of time that their identities will be coupled -- Sca1 switching appears to encode an elementary form of epigenetic memory that endows individual cells with a transient functional identity. Since the rate of switching is slower than the rate of cell division this allows the formation of communities of cells that maintain the same characteristics though divisions, and are therefore able to adopt a temporarily stable functional phenotype. Yet, by allowing mixing between the communities on a feasible time-scale, Sca1 fluctuations also safeguard long-term cell-cell variability and ensure that a robustly heterogeneous population, able to rapidly respond to a range of environmental challenges and resilient to removal of cellular sub-types, is maintained.
\begin{figure}[t!]
\includegraphics[width=0.475\textwidth]{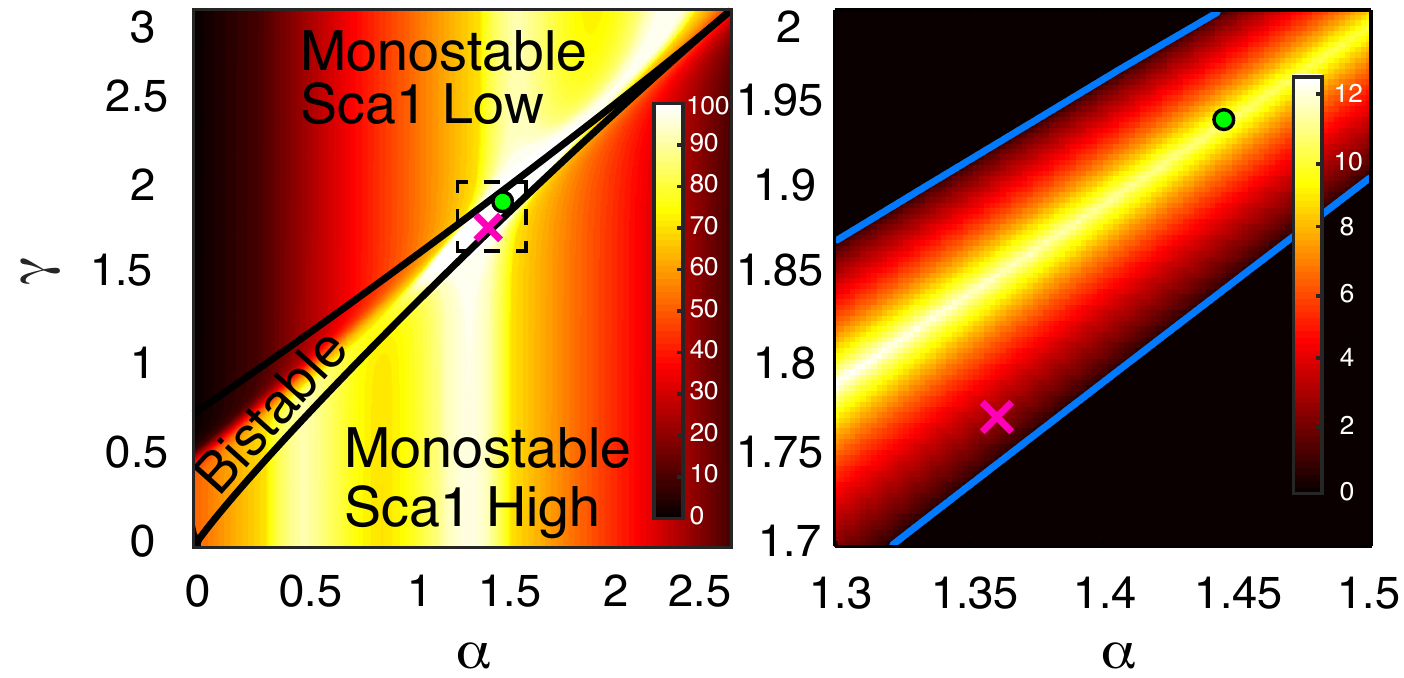}
\caption{(\emph{Left}) Entropy of the stationary distribution relative to the maximum entropy distribution over the $\alpha\gamma$-plane. The empirical distribution is marked with a magenta cross and the maximum entropy distribution $\rho^{\textrm{max}}_\infty (x)$ is marked with a green circle. Color shows percentiles. (\emph{Right}) Minimum \textsc{mfpt} $\tau$ in the vicinity of the maximum entropy distribution (close-up over the dashed box in the left panel). The critical lines separating the bistable and monostable regimes are shown in blue. The empirical distribution lies in the small region of the $\alpha\gamma$-plane that is both close to critical and of high entropy. Color shows dimensionless time. Color online.}
\label{FIG:WL}
\end{figure}

These results indicate that regulated fluctuations in Sca1 levels may be an intrinsic feature of \textsc{eml} cells in culture since they provide a mechanism by which the population hedges against unforeseen future environmental challenges and thereby retains the capacity to differentiate along both erythroid and/or myeloid lineages as required. If this is the case, then it is natural to ask if the experimentally observed stationary Sca1 distribution is optimal for this purpose; that is, if it is \emph{maximally} variable in some appropriately defined way. To investigate this, it is convenient to introduce a parameterization of the potential $\psi(x)$, in order to compare distributions. A parsimonious model, which allows for observed bimodality without introducing large numbers of parameters, is:
\begin{equation} 
\frac{\mathrm{d} \psi}{\mathrm{d} x} = \beta x - \alpha_0 - \frac{\alpha_1 x^n}{K^n + x^n}, \nonumber
\end{equation}
where $n$ is a positive even integer \footnote{This restriction ensures that $\psi(x)$ is continuous and real for all $x \in \mathbb{R}$. Although in principle $x$ may be negative, Sca1 levels are sufficiently high that we did not observe negative values in practice.}. Intuitively, this is a simple model of a positive-feedback based bistable switch of the kind that commonly regulate cell fate changes \cite{becskei2001positive, *xiong2003positive, *ferrell2002self}. The stationary distribution $p_\infty (x)$ is then characterized by four nonnegative dimensionless parameters: $\boldsymbol{\theta}$ = [$n$, $\alpha =\alpha_0 /\alpha_1$, $\gamma= \beta K /\alpha_1$, $\sigma_{\textrm{d}}=\sigma\beta/\alpha_1^2$]. 

For fixed $\boldsymbol{\theta}$, the conditional probability $\rho_\infty(x \, | \, \boldsymbol{\theta})$ is the minimizer of the free energy $F(\rho$), and may therefore be viewed as the most non-committal way to assign probabilities subject to the particular constrains imposed upon the dynamics by $\psi (x; \boldsymbol{\theta})$ (i.e. an instance of maximum entropy statistical inference) \cite{jaynes}. As each set of model parameters defines a different potential, which places different constraints upon the dynamics, we may therefore determine the extent to which Sca1 fluctuations optimize population diversity by assessing the proximity of the empirical stationary Sca1 distribution to the maximum entropy distribution $\rho^{\textrm{max}}_\infty (x) = \rho_\infty (x \, | \, \boldsymbol{\theta}^{*})$, where $S ( \rho_\infty(x \, | \, \boldsymbol{\theta}^*)) = \max_{\boldsymbol{\theta}} \, S ( \rho_\infty(x \, | \, \boldsymbol{\theta}))$. The relative entropy,
\begin{equation}
D(\rho_\infty \, || \, \rho^{\textrm{max}}_\infty) = \int \rho_\infty \log \left ( \frac{\rho_\infty}{\rho_\infty^\textrm{max}} \right ) \, \mathrm{d}x,  \nonumber
\end{equation} 
is a natural way to measure this proximity. Since the Hill coefficient $n$ is, informally, a measure of the sensitivity of the underlying switch to input stimulus, it primarily affects the curvature of the potential around the local minima $x_0$ (where present) and does not have a strong effect on the entropy. However, by governing a cusp bifurcation that determines whether the underlying switch is in a monostable or bistable state, $\alpha$ and $\gamma$ can affect the entropy of the stationary distribution considerably. Fig. \ref{FIG:WL} shows how the relative entropy of $p_\infty (x)$ varies over the biologically relevant bistable region of the $\alpha\gamma$-plane \footnote{Note that $p_\infty (x)$ also depends upon $\sigma_{\textrm{d}}$, the relative strength of stochastic fluctuations. However, since this parameter includes the effects of unregulated extrinsic noise, we assume that it is not within the cells capacity to regulate and fix it at the experimentally determined value}. It can be seen that the point estimate for the experimentally observed Sca1 distribution is remarkably close to the maximum entropy distribution $\rho^{\textrm{max}}_\infty (x)$. However, while the maximum entropy distribution is in the center of the bistable regime, the empirical distribution is close to one of the critical lines that separate the bistable and monostable regimes (shown in blue in Fig. \ref{FIG:WL}, right). It has long been suggested that such criticality may emerge naturally in biological systems via self-organizing evolutionary processes without the need for fine-tuning (i.e. as an attractor of the evolutionary dynamics) since critical states provide the dual benefits of stability and adaptability \cite{pal2015non, *nykter2008gene, *stuart1993origins}. Here, proximity to criticality specifically regulates the rate of mixing between the Sca1 high and low subpopulations, and therefore the response time of the population to environmental changes. To illustrate this, Fig. 3 also shows how $\tau = \min \, [ \tau_-, \tau_+ ]$, where $\tau_-$ and $\tau_+$ are the mean first passage times (\textsc{mfpt}s) in the low and high states respectively, varies in the vicinity of the maximum entropy state in the $\alpha\gamma$-plane. It can be seen that the minimum \textsc{mfpt} in the maximum entropy state is approximately an order of magnitude greater than that of the empirical distribution. Thus, while a population distributed according to the maximum entropy distribution would ultimately able to adapt better to environmental changes than the empirical population, it could not do so as rapidly. In this regard, close proximity to criticality is vital since it ensures that a diverse population is produced, yet mixing between subpopulations occurs on a physically relevant time-scale. These results suggest that Sca1 levels are regulated by fitness constraints that involve a trade-off between maximizing cell-cell variability and maintaining the ability to respond rapidly to environmental changes.

In summary, we have proposed an information-theoretic interpretation of stem cell dynamics that views cellular multipotency as an instance of maximum entropy statistical inference. Although we have focused on Sca1 dynamics, comperable expression fluctuations are known to generate functional diversity in other mammalian stem cell systems \cite{chambers2007nanog,*hayashi2008dynamic,*toyooka2008identification,*kalmar2009regulated,*kobayashi2009cyclic,*canham2010functional,*macfarlan2012embryonic,*trott2012dissecting}, and similar ergodic dynamics have been observed to give rise to universal protein expression distributions in microorganisms \cite{salman2012universal, *brenner2015single}. 
Thus, the generation of ergodic expression fluctuations may be a generic way in which cell populations maintain robust multilineage differentiation potential under environmental uncertainty. If so, then molecular noise processing could be particularly important in regulating stem cell function in a range of contexts. A better understanding of the relationship between molecular noise and stem cell identity should help to distinguish variability due to inter-changeable subpopulations of cells from that due to the presence of distinct, non-interconvertible, cell types (i.e. to determine which underlying stochastic processes are ergodic) \cite{pina2012inferring, *rue2015cell}. We anticipate that advances in single cell profiling techniques will help to address these issues in the near future.

\textbf{Acknowledgements}. The authors would like to thank Sui Huang for providing the experimental data. This work is partly funded by BBSRC grant BB/L000512/1.

\bibliographystyle{apsrev4-1}

\newpage
\beginsupplement
\onecolumngrid
\appendix

\section{Supplementary Material}
We assume that the intracellular dynamics of Sca1 expression level $z(t)$ are given by the following generic stochastic differential equation:
\begin{equation} \label{sde}
\frac{\textrm{d}z}{\textrm{d}t}=a(z)+\sqrt{2d(z)}\xi(t), \nonumber
\end{equation}
where $\xi(t)$ is a standard one-dimensional white noise process [$\left \langle \xi(t) \right \rangle=0$ and $\left \langle \xi(t)\xi(s) \right \rangle =\delta(t-s)$] and $d(z)$ accounts for fluctuations in Sca1 levels due to both intrinsic and extrinsic sources. This equation can be written in a more convenient form by introducing an appropriate reaction coordinate $x(z)$ such that the dynamics are mapped to those of a Brownian particle in a one-dimensional potential field. Such a transformation may be achieved by application of It\={o}'s lemma, which reads:
\begin{equation}
\frac{\textrm{d}x}{\textrm{d}t} = a(z) \frac{\textrm{d}x}{\textrm{d}z} + d(z) \frac{\textrm{d}^2x}{\textrm{d}z^2}  + \sqrt{2d(z)}\frac{\textrm{d}x}{\textrm{d}z} \xi(t). \nonumber
\end{equation}
Firstly, the reaction coordinate $x(z)$ can be chosen such that the noise term in this equation is constant, say $\sqrt{2\sigma}$, which gives the transformation
\begin{equation}
x = \int \sqrt{\frac{\sigma}{d(z)}} \textrm{d}z. \label{fzt}
\end{equation}
Since the dynamics are one-dimensional we may also introduce a potential $\psi(x)$ such that
\begin{equation}
-\frac{\textrm{d}\psi}{\textrm{d}x} = a(z) \frac{\textrm{d}x}{\textrm{d}z} + d(z) \frac{\textrm{d}^2x}{\textrm{d}z^2}, \nonumber
\end{equation}
to obtain
\begin{equation}
\frac{\textrm{d}x}{\textrm{d}t}=-\frac{\textrm{d}\psi}{\textrm{d}x} + \sqrt{2\sigma} \xi(t), \nonumber
\end{equation}
which is the stochastic differential equation corresponding to the Fokker-Planck equation given in the main text. Experimental data suggests that protein expression fluctuations often scale linearly with expression level \cite{Bar-Even}. Thus, a natural choice for the noise term is $d(z) = \sigma z^2$. Substituting this into Eq.~\eqref{fzt} gives $x= \log(z)$. This approach is similar to that taken in Ref. \cite{sisan2012predicting}. 

\bibliographystyle{apsrev4-1}

\begin{thebibliography}{39}%
\makeatletter
\providecommand \@ifxundefined [1]{%
 \@ifx{#1\undefined}
}%
\providecommand \@ifnum [1]{%
 \ifnum #1\expandafter \@firstoftwo
 \else \expandafter \@secondoftwo
 \fi
}%
\providecommand \@ifx [1]{%
 \ifx #1\expandafter \@firstoftwo
 \else \expandafter \@secondoftwo
 \fi
}%
\providecommand \natexlab [1]{#1}%
\providecommand \enquote  [1]{``#1''}%
\providecommand \bibnamefont  [1]{#1}%
\providecommand \bibfnamefont [1]{#1}%
\providecommand \citenamefont [1]{#1}%
\providecommand \href@noop [0]{\@secondoftwo}%
\providecommand \href [0]{\begingroup \@sanitize@url \@href}%
\providecommand \@href[1]{\@@startlink{#1}\@@href}%
\providecommand \@@href[1]{\endgroup#1\@@endlink}%
\providecommand \@sanitize@url [0]{\catcode `\\12\catcode `\$12\catcode
  `\&12\catcode `\#12\catcode `\^12\catcode `\_12\catcode `\%12\relax}%
\providecommand \@@startlink[1]{}%
\providecommand \@@endlink[0]{}%
\providecommand \url  [0]{\begingroup\@sanitize@url \@url }%
\providecommand \@url [1]{\endgroup\@href {#1}{\urlprefix }}%
\providecommand \urlprefix  [0]{URL }%
\providecommand \Eprint [0]{\href }%
\providecommand \doibase [0]{http://dx.doi.org/}%
\providecommand \selectlanguage [0]{\@gobble}%
\providecommand \bibinfo  [0]{\@secondoftwo}%
\providecommand \bibfield  [0]{\@secondoftwo}%
\providecommand \translation [1]{[#1]}%
\providecommand \BibitemOpen [0]{}%
\providecommand \bibitemStop [0]{}%
\providecommand \bibitemNoStop [0]{.\EOS\space}%
\providecommand \EOS [0]{\spacefactor3000\relax}%
\providecommand \BibitemShut  [1]{\csname bibitem#1\endcsname}%
\let\auto@bib@innerbib\@empty
\bibitem [{\citenamefont {Veening}\ \emph
  {et~al.}(2008{\natexlab{a}})\citenamefont {Veening}, \citenamefont {Stewart},
  \citenamefont {Berngruber}, \citenamefont {Taddei}, \citenamefont {Kuipers},\
  and\ \citenamefont {Hamoen}}]{Veening}%
  \BibitemOpen
  \bibfield  {author} {\bibinfo {author} {\bibfnamefont {J.~W.}\ \bibnamefont
  {Veening}}, \bibinfo {author} {\bibfnamefont {E.~J.}\ \bibnamefont
  {Stewart}}, \bibinfo {author} {\bibfnamefont {T.~W.}\ \bibnamefont
  {Berngruber}}, \bibinfo {author} {\bibfnamefont {F.}~\bibnamefont {Taddei}},
  \bibinfo {author} {\bibfnamefont {O.~P.}\ \bibnamefont {Kuipers}}, \ and\
  \bibinfo {author} {\bibfnamefont {L.~W.}\ \bibnamefont {Hamoen}},\
  }\href@noop {} {\bibfield  {journal} {\bibinfo  {journal} {Proc. Natl. Acad.
  Sci. USA}\ }\textbf {\bibinfo {volume} {105}},\ \bibinfo {pages} {4393}
  (\bibinfo {year} {2008}{\natexlab{a}})}\BibitemShut {NoStop}%
\bibitem [{\citenamefont {Veening}\ \emph
  {et~al.}(2008{\natexlab{b}})\citenamefont {Veening}, \citenamefont {Smits},\
  and\ \citenamefont {Kuipers}}]{Veening2}%
  \BibitemOpen
  \bibfield  {author} {\bibinfo {author} {\bibfnamefont {J.~W.}\ \bibnamefont
  {Veening}}, \bibinfo {author} {\bibfnamefont {W.~K.}\ \bibnamefont {Smits}},
  \ and\ \bibinfo {author} {\bibfnamefont {O.~P.}\ \bibnamefont {Kuipers}},\
  }\href@noop {} {\bibfield  {journal} {\bibinfo  {journal} {Annu. Rev.
  Microbiol.}\ }\textbf {\bibinfo {volume} {62}},\ \bibinfo {pages} {193}
  (\bibinfo {year} {2008}{\natexlab{b}})}\BibitemShut {NoStop}%
\bibitem [{\citenamefont {Graf}\ and\ \citenamefont {Stadtfeld}(2008)}]{Graf}%
  \BibitemOpen
  \bibfield  {author} {\bibinfo {author} {\bibfnamefont {T.}~\bibnamefont
  {Graf}}\ and\ \bibinfo {author} {\bibfnamefont {M.}~\bibnamefont
  {Stadtfeld}},\ }\href@noop {} {\bibfield  {journal} {\bibinfo  {journal}
  {Cell Stem Cell}\ }\textbf {\bibinfo {volume} {3}},\ \bibinfo {pages} {480}
  (\bibinfo {year} {2008})}\BibitemShut {NoStop}%
\bibitem [{\citenamefont {Cahan}\ and\ \citenamefont {Daley}(2013)}]{Cahan}%
  \BibitemOpen
  \bibfield  {author} {\bibinfo {author} {\bibfnamefont {P.}~\bibnamefont
  {Cahan}}\ and\ \bibinfo {author} {\bibfnamefont {G.~Q.}\ \bibnamefont
  {Daley}},\ }\href@noop {} {\bibfield  {journal} {\bibinfo  {journal} {Nat.
  Rev. Mol. Cell. Biol.}\ }\textbf {\bibinfo {volume} {14}},\ \bibinfo {pages}
  {357} (\bibinfo {year} {2013})}\BibitemShut {NoStop}%
\bibitem [{\citenamefont {Kalmar}\ \emph {et~al.}(2009)\citenamefont {Kalmar},
  \citenamefont {Lim}, \citenamefont {Hayward}, \citenamefont
  {Mu{\~n}oz-Descalzo}, \citenamefont {Nichols}, \citenamefont
  {Garcia-Ojalvo},\ and\ \citenamefont {Arias}}]{kalmar2009regulated}%
  \BibitemOpen
  \bibfield  {author} {\bibinfo {author} {\bibfnamefont {T.}~\bibnamefont
  {Kalmar}}, \bibinfo {author} {\bibfnamefont {C.}~\bibnamefont {Lim}},
  \bibinfo {author} {\bibfnamefont {P.}~\bibnamefont {Hayward}}, \bibinfo
  {author} {\bibfnamefont {S.}~\bibnamefont {Mu{\~n}oz-Descalzo}}, \bibinfo
  {author} {\bibfnamefont {J.}~\bibnamefont {Nichols}}, \bibinfo {author}
  {\bibfnamefont {J.}~\bibnamefont {Garcia-Ojalvo}}, \ and\ \bibinfo {author}
  {\bibfnamefont {A.~M.}\ \bibnamefont {Arias}},\ }\href@noop {} {\bibfield
  {journal} {\bibinfo  {journal} {PLoS Biol.}\ }\textbf {\bibinfo {volume}
  {7}},\ \bibinfo {pages} {e1000149} (\bibinfo {year} {2009})}\BibitemShut
  {NoStop}%
\bibitem [{\citenamefont {Lei}\ \emph {et~al.}(2014)\citenamefont {Lei},
  \citenamefont {Levin},\ and\ \citenamefont {Nie}}]{lei}%
  \BibitemOpen
  \bibfield  {author} {\bibinfo {author} {\bibfnamefont {J.}~\bibnamefont
  {Lei}}, \bibinfo {author} {\bibfnamefont {S.}~\bibnamefont {Levin}}, \ and\
  \bibinfo {author} {\bibfnamefont {Q.}~\bibnamefont {Nie}},\ }\href@noop {}
  {\bibfield  {journal} {\bibinfo  {journal} {Proc. Natl. Acad. Sci. USA}\
  }\textbf {\bibinfo {volume} {111}},\ \bibinfo {pages} {E880} (\bibinfo {year}
  {2014})}\BibitemShut {NoStop}%
\bibitem [{\citenamefont {MacArthur}\ and\ \citenamefont
  {Lemischka}(2013)}]{MacArthur}%
  \BibitemOpen
  \bibfield  {author} {\bibinfo {author} {\bibfnamefont {B.}~\bibnamefont
  {MacArthur}}\ and\ \bibinfo {author} {\bibfnamefont {I.~R.}\ \bibnamefont
  {Lemischka}},\ }\href@noop {} {\bibfield  {journal} {\bibinfo  {journal}
  {Cell}\ }\textbf {\bibinfo {volume} {154}},\ \bibinfo {pages} {484} (\bibinfo
  {year} {2013})}\BibitemShut {NoStop}%
\bibitem [{\citenamefont {Garcia-Ojalvo}\ and\ \citenamefont
  {Arias}(2012)}]{garcia2012towards}%
  \BibitemOpen
  \bibfield  {author} {\bibinfo {author} {\bibfnamefont {J.}~\bibnamefont
  {Garcia-Ojalvo}}\ and\ \bibinfo {author} {\bibfnamefont {A.~M.}\ \bibnamefont
  {Arias}},\ }\href@noop {} {\bibfield  {journal} {\bibinfo  {journal} {Curr.
  Opin. Genet. Dev.}\ }\textbf {\bibinfo {volume} {22}},\ \bibinfo {pages}
  {619} (\bibinfo {year} {2012})}\BibitemShut {NoStop}%
\bibitem [{\citenamefont {Osawa}\ \emph {et~al.}(1996)\citenamefont {Osawa},
  \citenamefont {Hanada}, \citenamefont {Hamada},\ and\ \citenamefont
  {Nakauchi}}]{Osawa}%
  \BibitemOpen
  \bibfield  {author} {\bibinfo {author} {\bibfnamefont {M.}~\bibnamefont
  {Osawa}}, \bibinfo {author} {\bibfnamefont {K.~I.}\ \bibnamefont {Hanada}},
  \bibinfo {author} {\bibfnamefont {H.}~\bibnamefont {Hamada}}, \ and\ \bibinfo
  {author} {\bibfnamefont {H.}~\bibnamefont {Nakauchi}},\ }\href@noop {}
  {\bibfield  {journal} {\bibinfo  {journal} {Science}\ }\textbf {\bibinfo
  {volume} {273}},\ \bibinfo {pages} {242} (\bibinfo {year}
  {1996})}\BibitemShut {NoStop}%
\bibitem [{\citenamefont {Wang}\ and\ \citenamefont {Jaenisch}(2004)}]{wang}%
  \BibitemOpen
  \bibfield  {author} {\bibinfo {author} {\bibfnamefont {Z.}~\bibnamefont
  {Wang}}\ and\ \bibinfo {author} {\bibfnamefont {R.}~\bibnamefont
  {Jaenisch}},\ }\href@noop {} {\bibfield  {journal} {\bibinfo  {journal} {Dev.
  Biol.}\ }\textbf {\bibinfo {volume} {275}},\ \bibinfo {pages} {192} (\bibinfo
  {year} {2004})}\BibitemShut {NoStop}%
\bibitem [{\citenamefont {Huang}\ \emph {et~al.}(2005)\citenamefont {Huang},
  \citenamefont {Eichler}, \citenamefont {Bar-Yam},\ and\ \citenamefont
  {Ingber}}]{huang2005cell}%
  \BibitemOpen
  \bibfield  {author} {\bibinfo {author} {\bibfnamefont {S.}~\bibnamefont
  {Huang}}, \bibinfo {author} {\bibfnamefont {G.}~\bibnamefont {Eichler}},
  \bibinfo {author} {\bibfnamefont {Y.}~\bibnamefont {Bar-Yam}}, \ and\
  \bibinfo {author} {\bibfnamefont {D.}~\bibnamefont {Ingber}},\ }\href@noop {}
  {\bibfield  {journal} {\bibinfo  {journal} {Phys. Rev. Lett.}\ }\textbf
  {\bibinfo {volume} {94}},\ \bibinfo {pages} {128701} (\bibinfo {year}
  {2005})}\BibitemShut {NoStop}%
\bibitem [{\citenamefont {Chang}\ \emph {et~al.}(2008)\citenamefont {Chang},
  \citenamefont {Hemberg}, \citenamefont {Barahona}, \citenamefont {Ingber},\
  and\ \citenamefont {Huang}}]{chang}%
  \BibitemOpen
  \bibfield  {author} {\bibinfo {author} {\bibfnamefont {H.}~\bibnamefont
  {Chang}}, \bibinfo {author} {\bibfnamefont {M.}~\bibnamefont {Hemberg}},
  \bibinfo {author} {\bibfnamefont {M.}~\bibnamefont {Barahona}}, \bibinfo
  {author} {\bibfnamefont {D.}~\bibnamefont {Ingber}}, \ and\ \bibinfo {author}
  {\bibfnamefont {S.}~\bibnamefont {Huang}},\ }\href@noop {} {\bibfield
  {journal} {\bibinfo  {journal} {Nature}\ }\textbf {\bibinfo {volume} {453}},\
  \bibinfo {pages} {544} (\bibinfo {year} {2008})}\BibitemShut {NoStop}%
\bibitem [{\citenamefont {Bar-Even}\ \emph {et~al.}(2006)\citenamefont
  {Bar-Even}, \citenamefont {Paulsson}, \citenamefont {Maheshri}, \citenamefont
  {Carmi}, \citenamefont {E.}, \citenamefont {Pilpel},\ and\ \citenamefont
  {Barkai}}]{Bar-Even}%
  \BibitemOpen
  \bibfield  {author} {\bibinfo {author} {\bibfnamefont {A.}~\bibnamefont
  {Bar-Even}}, \bibinfo {author} {\bibfnamefont {J.}~\bibnamefont {Paulsson}},
  \bibinfo {author} {\bibfnamefont {N.}~\bibnamefont {Maheshri}}, \bibinfo
  {author} {\bibfnamefont {M.}~\bibnamefont {Carmi}}, \bibinfo {author}
  {\bibfnamefont {O.}~\bibnamefont {E.}}, \bibinfo {author} {\bibfnamefont
  {Y.}~\bibnamefont {Pilpel}}, \ and\ \bibinfo {author} {\bibfnamefont
  {N.}~\bibnamefont {Barkai}},\ }\href@noop {} {\bibfield  {journal} {\bibinfo
  {journal} {Nat. Genet.}\ }\textbf {\bibinfo {volume} {38}},\ \bibinfo {pages}
  {636} (\bibinfo {year} {2006})}\BibitemShut {NoStop}%
\bibitem [{\citenamefont {Sisan}\ \emph {et~al.}(2012)\citenamefont {Sisan},
  \citenamefont {Halter}, \citenamefont {Hubbard},\ and\ \citenamefont
  {Plant}}]{sisan2012predicting}%
  \BibitemOpen
  \bibfield  {author} {\bibinfo {author} {\bibfnamefont {D.~R.}\ \bibnamefont
  {Sisan}}, \bibinfo {author} {\bibfnamefont {M.}~\bibnamefont {Halter}},
  \bibinfo {author} {\bibfnamefont {J.~B.}\ \bibnamefont {Hubbard}}, \ and\
  \bibinfo {author} {\bibfnamefont {A.~L.}\ \bibnamefont {Plant}},\ }\href@noop
  {} {\bibfield  {journal} {\bibinfo  {journal} {Proc. Natl. Acad. Sci. USA}\
  }\textbf {\bibinfo {volume} {109}},\ \bibinfo {pages} {19262} (\bibinfo
  {year} {2012})}\BibitemShut {NoStop}%
\bibitem [{\citenamefont {H{\"a}nggi}\ \emph {et~al.}(1990)\citenamefont
  {H{\"a}nggi}, \citenamefont {Talkner},\ and\ \citenamefont
  {Borkovec}}]{hanggi1990reaction}%
  \BibitemOpen
  \bibfield  {author} {\bibinfo {author} {\bibfnamefont {P.}~\bibnamefont
  {H{\"a}nggi}}, \bibinfo {author} {\bibfnamefont {P.}~\bibnamefont {Talkner}},
  \ and\ \bibinfo {author} {\bibfnamefont {M.}~\bibnamefont {Borkovec}},\
  }\href@noop {} {\bibfield  {journal} {\bibinfo  {journal} {Rev. Mod. Phys.}\
  }\textbf {\bibinfo {volume} {62}},\ \bibinfo {pages} {251} (\bibinfo {year}
  {1990})}\BibitemShut {NoStop}%
\bibitem [{\citenamefont {Markowich}\ and\ \citenamefont
  {Villani}(2000)}]{markowich2000trend}%
  \BibitemOpen
  \bibfield  {author} {\bibinfo {author} {\bibfnamefont {P.~A.}\ \bibnamefont
  {Markowich}}\ and\ \bibinfo {author} {\bibfnamefont {C.}~\bibnamefont
  {Villani}},\ }\href@noop {} {\bibfield  {journal} {\bibinfo  {journal} {Mat.
  Contemp.}\ }\textbf {\bibinfo {volume} {19}},\ \bibinfo {pages} {1} (\bibinfo
  {year} {2000})}\BibitemShut {NoStop}%
\bibitem [{\citenamefont {Jordan}\ \emph {et~al.}(1998)\citenamefont {Jordan},
  \citenamefont {Kinderlehrer},\ and\ \citenamefont
  {Otto}}]{jordan1998variational}%
  \BibitemOpen
  \bibfield  {author} {\bibinfo {author} {\bibfnamefont {R.}~\bibnamefont
  {Jordan}}, \bibinfo {author} {\bibfnamefont {D.}~\bibnamefont
  {Kinderlehrer}}, \ and\ \bibinfo {author} {\bibfnamefont {F.}~\bibnamefont
  {Otto}},\ }\href@noop {} {\bibfield  {journal} {\bibinfo  {journal} {SIAM J.
  Math. Anal.}\ }\textbf {\bibinfo {volume} {29}},\ \bibinfo {pages} {1}
  (\bibinfo {year} {1998})}\BibitemShut {NoStop}%
\bibitem [{\citenamefont {Bowie}\ \emph {et~al.}(2006)\citenamefont {Bowie},
  \citenamefont {McKnight}, \citenamefont {Kent}, \citenamefont {McCaffrey},
  \citenamefont {Hoodless}, \citenamefont {Eaves} \emph {et~al.}}]{bowie}%
  \BibitemOpen
  \bibfield  {author} {\bibinfo {author} {\bibfnamefont {M.~B.}\ \bibnamefont
  {Bowie}}, \bibinfo {author} {\bibfnamefont {K.~D.}\ \bibnamefont {McKnight}},
  \bibinfo {author} {\bibfnamefont {D.~G.}\ \bibnamefont {Kent}}, \bibinfo
  {author} {\bibfnamefont {L.}~\bibnamefont {McCaffrey}}, \bibinfo {author}
  {\bibfnamefont {P.~A.}\ \bibnamefont {Hoodless}}, \bibinfo {author}
  {\bibfnamefont {C.~J.}\ \bibnamefont {Eaves}},  \emph {et~al.},\ }\href@noop
  {} {\bibfield  {journal} {\bibinfo  {journal} {J. Clin. Invest.}\ }\textbf
  {\bibinfo {volume} {116}},\ \bibinfo {pages} {2808} (\bibinfo {year}
  {2006})}\BibitemShut {NoStop}%
\bibitem [{\citenamefont {Nygren}\ \emph {et~al.}(2006)\citenamefont {Nygren},
  \citenamefont {Bryder},\ and\ \citenamefont {Jacobsen}}]{nygren}%
  \BibitemOpen
  \bibfield  {author} {\bibinfo {author} {\bibfnamefont {J.~M.}\ \bibnamefont
  {Nygren}}, \bibinfo {author} {\bibfnamefont {D.}~\bibnamefont {Bryder}}, \
  and\ \bibinfo {author} {\bibfnamefont {S.~E.~W.}\ \bibnamefont {Jacobsen}},\
  }\href@noop {} {\bibfield  {journal} {\bibinfo  {journal} {J. Immunol.}\
  }\textbf {\bibinfo {volume} {177}},\ \bibinfo {pages} {201} (\bibinfo {year}
  {2006})}\BibitemShut {NoStop}%
\bibitem [{Note1()}]{Note1}%
  \BibitemOpen
  \bibinfo {note} {This restriction ensures that $\psi (x)$ is continuous and
  real for all $x \in \protect \mathbb {R}$. Although in principle $x$ may be
  negative, Sca1 levels are sufficiently high that we did not observe negative
  values in practice.}\BibitemShut {Stop}%
\bibitem [{\citenamefont {Becskei}\ \emph {et~al.}(2001)\citenamefont
  {Becskei}, \citenamefont {S{\'e}raphin},\ and\ \citenamefont
  {Serrano}}]{becskei2001positive}%
  \BibitemOpen
  \bibfield  {author} {\bibinfo {author} {\bibfnamefont {A.}~\bibnamefont
  {Becskei}}, \bibinfo {author} {\bibfnamefont {B.}~\bibnamefont
  {S{\'e}raphin}}, \ and\ \bibinfo {author} {\bibfnamefont {L.}~\bibnamefont
  {Serrano}},\ }\href@noop {} {\bibfield  {journal} {\bibinfo  {journal} {EMBO
  J.}\ }\textbf {\bibinfo {volume} {20}},\ \bibinfo {pages} {2528} (\bibinfo
  {year} {2001})}\BibitemShut {NoStop}%
\bibitem [{\citenamefont {Xiong}\ and\ \citenamefont
  {Ferrell}(2003)}]{xiong2003positive}%
  \BibitemOpen
  \bibfield  {author} {\bibinfo {author} {\bibfnamefont {W.}~\bibnamefont
  {Xiong}}\ and\ \bibinfo {author} {\bibfnamefont {J.~E.}\ \bibnamefont
  {Ferrell}},\ }\href@noop {} {\bibfield  {journal} {\bibinfo  {journal}
  {Nature}\ }\textbf {\bibinfo {volume} {426}},\ \bibinfo {pages} {460}
  (\bibinfo {year} {2003})}\BibitemShut {NoStop}%
\bibitem [{\citenamefont {Ferrell}(2002)}]{ferrell2002self}%
  \BibitemOpen
  \bibfield  {author} {\bibinfo {author} {\bibfnamefont {J.~E.}\ \bibnamefont
  {Ferrell}},\ }\href@noop {} {\bibfield  {journal} {\bibinfo  {journal} {Curr.
  Opin. Cell. Biol.}\ }\textbf {\bibinfo {volume} {14}},\ \bibinfo {pages}
  {140} (\bibinfo {year} {2002})}\BibitemShut {NoStop}%
\bibitem [{\citenamefont {Jaynes}(1957)}]{jaynes}%
  \BibitemOpen
  \bibfield  {author} {\bibinfo {author} {\bibfnamefont {E.~T.}\ \bibnamefont
  {Jaynes}},\ }\href@noop {} {\bibfield  {journal} {\bibinfo  {journal} {Phys.
  Rev.}\ }\textbf {\bibinfo {volume} {106}},\ \bibinfo {pages} {620} (\bibinfo
  {year} {1957})}\BibitemShut {NoStop}%
\bibitem [{Note2()}]{Note2}%
  \BibitemOpen
  \bibinfo {note} {Note that $p_\infty (x)$ also depends upon $\sigma
  _{\protect \textrm {d}}$, the relative strength of stochastic fluctuations.
  However, since this parameter includes the effects of unregulated extrinsic
  noise, we assume that it is not within the cells capacity to regulate and fix
  it at the experimentally determined value}\BibitemShut {NoStop}%
\bibitem [{\citenamefont {Pal}\ \emph {et~al.}(2015)\citenamefont {Pal},
  \citenamefont {Ghosh},\ and\ \citenamefont {Bose}}]{pal2015non}%
  \BibitemOpen
  \bibfield  {author} {\bibinfo {author} {\bibfnamefont {M.}~\bibnamefont
  {Pal}}, \bibinfo {author} {\bibfnamefont {S.}~\bibnamefont {Ghosh}}, \ and\
  \bibinfo {author} {\bibfnamefont {I.}~\bibnamefont {Bose}},\ }\href@noop {}
  {\bibfield  {journal} {\bibinfo  {journal} {Phys. Biol.}\ }\textbf {\bibinfo
  {volume} {12}},\ \bibinfo {pages} {016001} (\bibinfo {year}
  {2015})}\BibitemShut {NoStop}%
\bibitem [{\citenamefont {Nykter}\ \emph {et~al.}(2008)\citenamefont {Nykter},
  \citenamefont {Price}, \citenamefont {Aldana}, \citenamefont {Ramsey},
  \citenamefont {Kauffman}, \citenamefont {Hood}, \citenamefont {Yli-Harja},\
  and\ \citenamefont {Shmulevich}}]{nykter2008gene}%
  \BibitemOpen
  \bibfield  {author} {\bibinfo {author} {\bibfnamefont {M.}~\bibnamefont
  {Nykter}}, \bibinfo {author} {\bibfnamefont {N.~D.}\ \bibnamefont {Price}},
  \bibinfo {author} {\bibfnamefont {M.}~\bibnamefont {Aldana}}, \bibinfo
  {author} {\bibfnamefont {S.~A.}\ \bibnamefont {Ramsey}}, \bibinfo {author}
  {\bibfnamefont {S.~A.}\ \bibnamefont {Kauffman}}, \bibinfo {author}
  {\bibfnamefont {L.~E.}\ \bibnamefont {Hood}}, \bibinfo {author}
  {\bibfnamefont {O.}~\bibnamefont {Yli-Harja}}, \ and\ \bibinfo {author}
  {\bibfnamefont {I.}~\bibnamefont {Shmulevich}},\ }\href@noop {} {\bibfield
  {journal} {\bibinfo  {journal} {Proc. Natl. Acad. Sci. USA}\ }\textbf
  {\bibinfo {volume} {105}},\ \bibinfo {pages} {1897} (\bibinfo {year}
  {2008})}\BibitemShut {NoStop}%
\bibitem [{\citenamefont {Kauffman}(1993)}]{stuart1993origins}%
  \BibitemOpen
  \bibfield  {author} {\bibinfo {author} {\bibfnamefont {S.~A.}\ \bibnamefont
  {Kauffman}},\ }\href@noop {} {\emph {\bibinfo {title} {The origins of order:
  self-organization and selection in evolution}}}\ (\bibinfo  {publisher}
  {Oxford University Press},\ \bibinfo {year} {1993})\BibitemShut {NoStop}%
\bibitem [{\citenamefont {Chambers}\ \emph {et~al.}(2007)\citenamefont
  {Chambers}, \citenamefont {Silva}, \citenamefont {Colby}, \citenamefont
  {Nichols}, \citenamefont {Nijmeijer}, \citenamefont {Robertson},
  \citenamefont {Vrana}, \citenamefont {Jones}, \citenamefont {Grotewold},\
  and\ \citenamefont {Smith}}]{chambers2007nanog}%
  \BibitemOpen
  \bibfield  {author} {\bibinfo {author} {\bibfnamefont {I.}~\bibnamefont
  {Chambers}}, \bibinfo {author} {\bibfnamefont {J.}~\bibnamefont {Silva}},
  \bibinfo {author} {\bibfnamefont {D.}~\bibnamefont {Colby}}, \bibinfo
  {author} {\bibfnamefont {J.}~\bibnamefont {Nichols}}, \bibinfo {author}
  {\bibfnamefont {B.}~\bibnamefont {Nijmeijer}}, \bibinfo {author}
  {\bibfnamefont {M.}~\bibnamefont {Robertson}}, \bibinfo {author}
  {\bibfnamefont {J.}~\bibnamefont {Vrana}}, \bibinfo {author} {\bibfnamefont
  {K.}~\bibnamefont {Jones}}, \bibinfo {author} {\bibfnamefont
  {L.}~\bibnamefont {Grotewold}}, \ and\ \bibinfo {author} {\bibfnamefont
  {A.}~\bibnamefont {Smith}},\ }\href@noop {} {\bibfield  {journal} {\bibinfo
  {journal} {Nature}\ }\textbf {\bibinfo {volume} {450}},\ \bibinfo {pages}
  {1230} (\bibinfo {year} {2007})}\BibitemShut {NoStop}%
\bibitem [{\citenamefont {Hayashi}\ \emph {et~al.}(2008)\citenamefont
  {Hayashi}, \citenamefont {de~Sousa~Lopes}, \citenamefont {Tang},\ and\
  \citenamefont {Surani}}]{hayashi2008dynamic}%
  \BibitemOpen
  \bibfield  {author} {\bibinfo {author} {\bibfnamefont {K.}~\bibnamefont
  {Hayashi}}, \bibinfo {author} {\bibfnamefont {S.~M.~C.}\ \bibnamefont
  {de~Sousa~Lopes}}, \bibinfo {author} {\bibfnamefont {F.}~\bibnamefont
  {Tang}}, \ and\ \bibinfo {author} {\bibfnamefont {M.~A.}\ \bibnamefont
  {Surani}},\ }\href@noop {} {\bibfield  {journal} {\bibinfo  {journal} {Cell
  Stem Cell}\ }\textbf {\bibinfo {volume} {3}},\ \bibinfo {pages} {391}
  (\bibinfo {year} {2008})}\BibitemShut {NoStop}%
\bibitem [{\citenamefont {Toyooka}\ \emph {et~al.}(2008)\citenamefont
  {Toyooka}, \citenamefont {Shimosato}, \citenamefont {Murakami}, \citenamefont
  {Takahashi},\ and\ \citenamefont {Niwa}}]{toyooka2008identification}%
  \BibitemOpen
  \bibfield  {author} {\bibinfo {author} {\bibfnamefont {Y.}~\bibnamefont
  {Toyooka}}, \bibinfo {author} {\bibfnamefont {D.}~\bibnamefont {Shimosato}},
  \bibinfo {author} {\bibfnamefont {K.}~\bibnamefont {Murakami}}, \bibinfo
  {author} {\bibfnamefont {K.}~\bibnamefont {Takahashi}}, \ and\ \bibinfo
  {author} {\bibfnamefont {H.}~\bibnamefont {Niwa}},\ }\href@noop {} {\bibfield
   {journal} {\bibinfo  {journal} {Development}\ }\textbf {\bibinfo {volume}
  {135}},\ \bibinfo {pages} {909} (\bibinfo {year} {2008})}\BibitemShut
  {NoStop}%
\bibitem [{\citenamefont {Kobayashi}\ \emph {et~al.}(2009)\citenamefont
  {Kobayashi}, \citenamefont {Mizuno}, \citenamefont {Imayoshi}, \citenamefont
  {Furusawa}, \citenamefont {Shirahige},\ and\ \citenamefont
  {Kageyama}}]{kobayashi2009cyclic}%
  \BibitemOpen
  \bibfield  {author} {\bibinfo {author} {\bibfnamefont {T.}~\bibnamefont
  {Kobayashi}}, \bibinfo {author} {\bibfnamefont {H.}~\bibnamefont {Mizuno}},
  \bibinfo {author} {\bibfnamefont {I.}~\bibnamefont {Imayoshi}}, \bibinfo
  {author} {\bibfnamefont {C.}~\bibnamefont {Furusawa}}, \bibinfo {author}
  {\bibfnamefont {K.}~\bibnamefont {Shirahige}}, \ and\ \bibinfo {author}
  {\bibfnamefont {R.}~\bibnamefont {Kageyama}},\ }\href@noop {} {\bibfield
  {journal} {\bibinfo  {journal} {Genes Dev.}\ }\textbf {\bibinfo {volume}
  {23}},\ \bibinfo {pages} {1870} (\bibinfo {year} {2009})}\BibitemShut
  {NoStop}%
\bibitem [{\citenamefont {Canham}\ \emph {et~al.}(2010)\citenamefont {Canham},
  \citenamefont {Sharov}, \citenamefont {Ko},\ and\ \citenamefont
  {Brickman}}]{canham2010functional}%
  \BibitemOpen
  \bibfield  {author} {\bibinfo {author} {\bibfnamefont {M.~A.}\ \bibnamefont
  {Canham}}, \bibinfo {author} {\bibfnamefont {A.~A.}\ \bibnamefont {Sharov}},
  \bibinfo {author} {\bibfnamefont {M.}~\bibnamefont {Ko}}, \ and\ \bibinfo
  {author} {\bibfnamefont {J.~M.}\ \bibnamefont {Brickman}},\ }\href@noop {}
  {\bibfield  {journal} {\bibinfo  {journal} {PLoS Biol.}\ }\textbf {\bibinfo
  {volume} {8}},\ \bibinfo {pages} {e1000379} (\bibinfo {year}
  {2010})}\BibitemShut {NoStop}%
\bibitem [{\citenamefont {Macfarlan}\ \emph {et~al.}(2012)\citenamefont
  {Macfarlan}, \citenamefont {Gifford}, \citenamefont {Driscoll}, \citenamefont
  {Lettieri}, \citenamefont {Rowe}, \citenamefont {Bonanomi}, \citenamefont
  {Firth}, \citenamefont {Singer}, \citenamefont {Trono},\ and\ \citenamefont
  {Pfaff}}]{macfarlan2012embryonic}%
  \BibitemOpen
  \bibfield  {author} {\bibinfo {author} {\bibfnamefont {T.~S.}\ \bibnamefont
  {Macfarlan}}, \bibinfo {author} {\bibfnamefont {W.~D.}\ \bibnamefont
  {Gifford}}, \bibinfo {author} {\bibfnamefont {S.}~\bibnamefont {Driscoll}},
  \bibinfo {author} {\bibfnamefont {K.}~\bibnamefont {Lettieri}}, \bibinfo
  {author} {\bibfnamefont {H.~M.}\ \bibnamefont {Rowe}}, \bibinfo {author}
  {\bibfnamefont {D.}~\bibnamefont {Bonanomi}}, \bibinfo {author}
  {\bibfnamefont {A.}~\bibnamefont {Firth}}, \bibinfo {author} {\bibfnamefont
  {O.}~\bibnamefont {Singer}}, \bibinfo {author} {\bibfnamefont
  {D.}~\bibnamefont {Trono}}, \ and\ \bibinfo {author} {\bibfnamefont {S.~L.}\
  \bibnamefont {Pfaff}},\ }\href@noop {} {\bibfield  {journal} {\bibinfo
  {journal} {Nature}\ }\textbf {\bibinfo {volume} {487}},\ \bibinfo {pages}
  {57} (\bibinfo {year} {2012})}\BibitemShut {NoStop}%
\bibitem [{\citenamefont {Trott}\ \emph {et~al.}(2012)\citenamefont {Trott},
  \citenamefont {Hayashi}, \citenamefont {Surani}, \citenamefont {Babu},\ and\
  \citenamefont {Martinez-Arias}}]{trott2012dissecting}%
  \BibitemOpen
  \bibfield  {author} {\bibinfo {author} {\bibfnamefont {J.}~\bibnamefont
  {Trott}}, \bibinfo {author} {\bibfnamefont {K.}~\bibnamefont {Hayashi}},
  \bibinfo {author} {\bibfnamefont {A.}~\bibnamefont {Surani}}, \bibinfo
  {author} {\bibfnamefont {M.~M.}\ \bibnamefont {Babu}}, \ and\ \bibinfo
  {author} {\bibfnamefont {A.}~\bibnamefont {Martinez-Arias}},\ }\href@noop {}
  {\bibfield  {journal} {\bibinfo  {journal} {Mol. Biosyst.}\ }\textbf
  {\bibinfo {volume} {8}},\ \bibinfo {pages} {744} (\bibinfo {year}
  {2012})}\BibitemShut {NoStop}%
\bibitem [{\citenamefont {Salman}\ \emph {et~al.}(2012)\citenamefont {Salman},
  \citenamefont {Brenner}, \citenamefont {Tung}, \citenamefont {Elyahu},
  \citenamefont {Stolovicki}, \citenamefont {Moore}, \citenamefont
  {Libchaber},\ and\ \citenamefont {Braun}}]{salman2012universal}%
  \BibitemOpen
  \bibfield  {author} {\bibinfo {author} {\bibfnamefont {H.}~\bibnamefont
  {Salman}}, \bibinfo {author} {\bibfnamefont {N.}~\bibnamefont {Brenner}},
  \bibinfo {author} {\bibfnamefont {C.-k.}\ \bibnamefont {Tung}}, \bibinfo
  {author} {\bibfnamefont {N.}~\bibnamefont {Elyahu}}, \bibinfo {author}
  {\bibfnamefont {E.}~\bibnamefont {Stolovicki}}, \bibinfo {author}
  {\bibfnamefont {L.}~\bibnamefont {Moore}}, \bibinfo {author} {\bibfnamefont
  {A.}~\bibnamefont {Libchaber}}, \ and\ \bibinfo {author} {\bibfnamefont
  {E.}~\bibnamefont {Braun}},\ }\href@noop {} {\bibfield  {journal} {\bibinfo
  {journal} {Phys. Rev. Lett.}\ }\textbf {\bibinfo {volume} {108}},\ \bibinfo
  {pages} {238105} (\bibinfo {year} {2012})}\BibitemShut {NoStop}%
\bibitem [{\citenamefont {Brenner}\ \emph {et~al.}(2015)\citenamefont
  {Brenner}, \citenamefont {Braun}, \citenamefont {Rotella},\ and\
  \citenamefont {Salman}}]{brenner2015single}%
  \BibitemOpen
  \bibfield  {author} {\bibinfo {author} {\bibfnamefont {N.}~\bibnamefont
  {Brenner}}, \bibinfo {author} {\bibfnamefont {E.}~\bibnamefont {Braun}},
  \bibinfo {author} {\bibfnamefont {J.~S.}\ \bibnamefont {Rotella}}, \ and\
  \bibinfo {author} {\bibfnamefont {H.}~\bibnamefont {Salman}},\ }\href@noop {}
  {\bibfield  {journal} {\bibinfo  {journal} {arXiv:1503.01046}\ } (\bibinfo
  {year} {2015})}\BibitemShut {NoStop}%
\bibitem [{\citenamefont {Pina}\ \emph {et~al.}(2012)\citenamefont {Pina},
  \citenamefont {Fugazza}, \citenamefont {Tipping}, \citenamefont {Brown},
  \citenamefont {Soneji}, \citenamefont {Teles}, \citenamefont {Peterson},\
  and\ \citenamefont {Enver}}]{pina2012inferring}%
  \BibitemOpen
  \bibfield  {author} {\bibinfo {author} {\bibfnamefont {C.}~\bibnamefont
  {Pina}}, \bibinfo {author} {\bibfnamefont {C.}~\bibnamefont {Fugazza}},
  \bibinfo {author} {\bibfnamefont {A.~J.}\ \bibnamefont {Tipping}}, \bibinfo
  {author} {\bibfnamefont {J.}~\bibnamefont {Brown}}, \bibinfo {author}
  {\bibfnamefont {S.}~\bibnamefont {Soneji}}, \bibinfo {author} {\bibfnamefont
  {J.}~\bibnamefont {Teles}}, \bibinfo {author} {\bibfnamefont
  {C.}~\bibnamefont {Peterson}}, \ and\ \bibinfo {author} {\bibfnamefont
  {T.}~\bibnamefont {Enver}},\ }\href@noop {} {\bibfield  {journal} {\bibinfo
  {journal} {Nat. Cell. Biol.}\ }\textbf {\bibinfo {volume} {14}},\ \bibinfo
  {pages} {287} (\bibinfo {year} {2012})}\BibitemShut {NoStop}%
\bibitem [{\citenamefont {Ru{\'e}}\ and\ \citenamefont
  {Martinez~Arias}(2015)}]{rue2015cell}%
  \BibitemOpen
  \bibfield  {author} {\bibinfo {author} {\bibfnamefont {P.}~\bibnamefont
  {Ru{\'e}}}\ and\ \bibinfo {author} {\bibfnamefont {A.}~\bibnamefont
  {Martinez~Arias}},\ }\href@noop {} {\bibfield  {journal} {\bibinfo  {journal}
  {Mol. Syst. Biol.}\ }\textbf {\bibinfo {volume} {11}} (\bibinfo {year}
  {2015})}\BibitemShut {NoStop}%
\end{thebibliography}

\begin{thebibliography}{2}%
\makeatletter
\providecommand \@ifxundefined [1]{%
 \@ifx{#1\undefined}
}%
\providecommand \@ifnum [1]{%
 \ifnum #1\expandafter \@firstoftwo
 \else \expandafter \@secondoftwo
 \fi
}%
\providecommand \@ifx [1]{%
 \ifx #1\expandafter \@firstoftwo
 \else \expandafter \@secondoftwo
 \fi
}%
\providecommand \natexlab [1]{#1}%
\providecommand \enquote  [1]{``#1''}%
\providecommand \bibnamefont  [1]{#1}%
\providecommand \bibfnamefont [1]{#1}%
\providecommand \citenamefont [1]{#1}%
\providecommand \href@noop [0]{\@secondoftwo}%
\providecommand \href [0]{\begingroup \@sanitize@url \@href}%
\providecommand \@href[1]{\@@startlink{#1}\@@href}%
\providecommand \@@href[1]{\endgroup#1\@@endlink}%
\providecommand \@sanitize@url [0]{\catcode `\\12\catcode `\$12\catcode
  `\&12\catcode `\#12\catcode `\^12\catcode `\_12\catcode `\%12\relax}%
\providecommand \@@startlink[1]{}%
\providecommand \@@endlink[0]{}%
\providecommand \url  [0]{\begingroup\@sanitize@url \@url }%
\providecommand \@url [1]{\endgroup\@href {#1}{\urlprefix }}%
\providecommand \urlprefix  [0]{URL }%
\providecommand \Eprint [0]{\href }%
\providecommand \doibase [0]{http://dx.doi.org/}%
\providecommand \selectlanguage [0]{\@gobble}%
\providecommand \bibinfo  [0]{\@secondoftwo}%
\providecommand \bibfield  [0]{\@secondoftwo}%
\providecommand \translation [1]{[#1]}%
\providecommand \BibitemOpen [0]{}%
\providecommand \bibitemStop [0]{}%
\providecommand \bibitemNoStop [0]{.\EOS\space}%
\providecommand \EOS [0]{\spacefactor3000\relax}%
\providecommand \BibitemShut  [1]{\csname bibitem#1\endcsname}%
\let\auto@bib@innerbib\@empty
\bibitem [{\citenamefont {Bar-Even}\ \emph {et~al.}(2006)\citenamefont
  {Bar-Even}, \citenamefont {Paulsson}, \citenamefont {Maheshri}, \citenamefont
  {Carmi}, \citenamefont {E.}, \citenamefont {Pilpel},\ and\ \citenamefont
  {Barkai}}]{Bar-Even}%
  \BibitemOpen
  \bibfield  {author} {\bibinfo {author} {\bibfnamefont {A.}~\bibnamefont
  {Bar-Even}}, \bibinfo {author} {\bibfnamefont {J.}~\bibnamefont {Paulsson}},
  \bibinfo {author} {\bibfnamefont {N.}~\bibnamefont {Maheshri}}, \bibinfo
  {author} {\bibfnamefont {M.}~\bibnamefont {Carmi}}, \bibinfo {author}
  {\bibfnamefont {O.}~\bibnamefont {E.}}, \bibinfo {author} {\bibfnamefont
  {Y.}~\bibnamefont {Pilpel}}, \ and\ \bibinfo {author} {\bibfnamefont
  {N.}~\bibnamefont {Barkai}},\ }\href@noop {} {\bibfield  {journal} {\bibinfo
  {journal} {Nat. Genet.}\ }\textbf {\bibinfo {volume} {38}},\ \bibinfo {pages}
  {636} (\bibinfo {year} {2006})}\BibitemShut {NoStop}%
\bibitem [{\citenamefont {Sisan}\ \emph {et~al.}(2012)\citenamefont {Sisan},
  \citenamefont {Halter}, \citenamefont {Hubbard},\ and\ \citenamefont
  {Plant}}]{sisan2012predicting}%
  \BibitemOpen
  \bibfield  {author} {\bibinfo {author} {\bibfnamefont {D.~R.}\ \bibnamefont
  {Sisan}}, \bibinfo {author} {\bibfnamefont {M.}~\bibnamefont {Halter}},
  \bibinfo {author} {\bibfnamefont {J.~B.}\ \bibnamefont {Hubbard}}, \ and\
  \bibinfo {author} {\bibfnamefont {A.~L.}\ \bibnamefont {Plant}},\ }\href@noop
  {} {\bibfield  {journal} {\bibinfo  {journal} {Proc. Natl. Acad. Sci. USA}\
  }\textbf {\bibinfo {volume} {109}},\ \bibinfo {pages} {19262} (\bibinfo
  {year} {2012})}\BibitemShut {NoStop}%
\end{thebibliography}

\end{document}